# A Medieval Multiverse: Mathematical Modelling of the 13th Century Universe of Robert Grosseteste

Richard G. Bower[1], Tom C. B. McLeish F.R.S.[1], Brian K. Tanner[1], Hannah E. Smithson[2], Cecilia Panti[3], Neil Lewis[4] and Giles E. M. Gasper[5]

[1]Department of Physics, Durham University, South Road, Durham DH1 3LE, U.K.
[2]Department of Experimental Psychology, Oxford University, South Parks Road, Oxford, OX1 3UD, U.K.
[3]Department of Business, Government and Philosophy, University of Rome-Tor Vergata, via Columbia 1, 00133 Roma, Italy
[4]Department of Philosophy, Georgetown University, Box 571133, New North 215, Washington, DC 20057-1133, U.S.A.
[5]Institute for Medieval and Early Modern Studies, Durham University, 7 Owengate, Durham, DH1 3HB, U.K.

Email: r.g.bower@durham.ac.uk, t.c.b.mcleish@dur.ac.uk

**Abstract**

In his treatise on light, written in about 1225, Robert Grosseteste describes a cosmological model in which the Universe is created in a big-bang like explosion and subsequent condensation. He postulates that the fundamental coupling of light and matter gives rises to the material body of the entire cosmos. Expansion is arrested when matter reaches a minimum density and subsequent emission of light from the outer region leads to compression and rarefaction of the inner bodily mass so as to create nine celestial spheres, with an imperfect residual core. In this paper we reformulate the Latin description in terms of a modern mathematical model. The equations which describe the coupling of light and matter are solved numerically, subject to initial conditions and critical criteria consistent with the text. Formation of a universe with a non-infinite number of perfected spheres is extremely sensitive to the initial conditions, the intensity of the light and the transparency of these spheres. In this "medieval multiverse", only a small range of opacity and initial density profiles lead to a stable universe with nine perfected spheres. As in current cosmological thinking, the existence of Grosseteste's universe relies on a very special combination of fundamental parameters.

**Introduction**

An early proponent of the newly rediscovered works of Aristotle, the methodology of Robert Grosseteste (ca. 1170-1253) was sufficiently revolutionary that some twentieth-century scholars claimed him as the first modern scientist and the antecessor of the scientific method. Among his scientific writings (1), his treatise on light (*De luce*) [2,3] is the most famous and extensively quoted, with (misplaced, but thought-provoking) claims [4] that he predicted the "Big Bang" theory of cosmological expansion eight centuries ahead of Lemaitre [5] and Hubble [6]. In this article, we express Grosseteste's model of how light interacts with matter in terms of modern mathematics and show that it can indeed generate his claimed structure of the Universe.

While it is crucial to avoid superposing a modern world view into Grosseteste's thought, throughout his work there pervades an interest in the nature of the created world, the existence of order within it, the mechanisms whereby it is sustained and a search for unity of explanation. These ideas are common in medieval thinking; nonetheless the originality of Grosseteste was to think about unity, order and causal explanation of natural phenomena as being due to light, its properties and the mechanism by which we perceive it. Because of his contribution to the development of scientific thought in the early 13$^{th}$ century, underpinned by belief in an ordered universe, Grosseteste has been the focus of collaboration between medievalists and scientists to examine his works using modern analytical techniques and methodology. Examination of Grosseteste's shorter treatise on colour *(De colore)* using palaeographic, linguistic, contextual and scientific reasoning has resulted in reinterpretation of Grosseteste's model of colour formation [7] and its explanation within the framework of modern colour theory [8]. We are not trying to "correct" Grosseteste's thinking in the light of modern physics, nor are we claiming Grosseteste's ideas as a precedent for modern cosmological thinking. Rather, we are making a translation, not just from Latin into English but from the new critical Latin edition [9] and English translation [10] of his *De luce* into mathematical language. We aim to write down the equations, as he might have done had he access to modern mathematical and computational techniques, solve the equations numerically and explore the solutions. There are benefits here from both an historical and a scientific perspective. The application of mathematics and computation generate, as we shall see, a closer and more comprehensive examination of a medieval scientific text and the mind behind it. However, there are scientific benefits as well, as the medieval cosmos constitutes a quite novel arena to compute radiation/matter interactions and dynamics, and in which to discover new physical structure.

**Grosseteste's model of light and its interaction with matter**

*De luce* was probably written about 1225, almost contemporaneously with *De colore,* although almost nothing is known of Grosseteste's whereabouts at this time in his life. Grosseteste began *De luce* by immediately making the bold postulates that light (*lux* in the Latin) is the first corporeal form and that it multiplies itself infinitely, expanding instantaneously from a point into a sphere of any size. He argued that neither the attribute of corporeal form nor matter has dimension but that, because form and matter are inseparable, by its expansion into all directions light introduces the three dimensions into matter. In the beginning of time, light extended matter, drawing it out along with itself into a sphere the size of the material universe.

It is this initial expansion of the universe which has attracted most attention but Grosseteste's really innovative intellectual construction was to link cosmology with his light-based conception of how ordinary material bodies become extended. He realised that as light drags matter outwards, and implicitly postulating conservation of matter, the density must decrease as the radius increases. As a vacuum is impossible within the Aristotelian framework, there must be a minimum density beyond which matter cannot be rarefied and this sets the boundary of the Universe. Grosseteste asserted that at this minimum density, there is a phase change (or perfection) of matter-plus-light and this perfect state can undergo no further change, forming the first celestial sphere of the cosmos. He then asserted that this perfect body, consisting only of first form (*lux*) and first matter, itself emits light of a different kind (*lumen* in the Latin) towards the centre of the sphere which is now the boundary of the universe. As it propagates, it sweeps up the (imperfect) matter, or body [10] (Latin *corpus*), compressing it. Because the first sphere is perfect and cannot change its status, and because there cannot be space that is empty, the *lumen* it emits sweeps up and compresses the matter inside the sphere as it propagates inwards. Grosseteste argued that the lumen propagated almost instantaneously towards the centre of the sphere, and in doing so rarefied matter in the outer region, which consequently became perfected and could no longer undergo change. The swept up material generates the second of the heavenly spheres, namely that of the fixed stars, corresponding to the first sphere of the Aristotelian cosmology. The light (*lux*) present in the first sphere is doubled in the second. Similarly, *lumen* is emitted from the second perfected sphere, sweeps up matter until there is further rarefaction and compression leading to a third perfected sphere. This continues until the ninth sphere, that of the moon, whose *lumen* emission is not sufficient to completely perfect the spheres of the elements (fire, air, water, earth) and these thus do not allow circular motion, which pertains only to perfect bodies, but just radial motion, and the latter two have the attribute of weight, due to their extremal density and compression.

**Mathematical statement of Grosseteste's model**

We will not focus on the initial expansion of the universe as Grosseteste's description was too brief to arrive at a particular model, specifying only the spherical symmetry of the distribution and the greater rarefaction of matter at greater radii. As there is at this point also nothing within the cosmos setting any special length-scale, it is reasonable to take a scaling (power-law) form for this distribution. We therefore write the initial condition for matter density in Grosseteste's universe as,

$$\rho(r;t=0) = \rho_{c0} r^{-\alpha} \qquad \textit{(for } r < r_{max}(t)\textit{)} \qquad [1]$$

The matter at the outer radius ($r_{max}$) is at the minimum density $\rho_{c0}$. The radius at which this happens determines the size of the universe. With no loss of generality, we will set our length scale so that this occurs at *r=1*. To describe how the formation of the first "perfected" matter sets off a chain of events that eventually leads to the formation of the terrestrial world, we must provide a mathematical description of how *lumen* propagates and formulate its interaction with matter.

We require a computable set of field equations for body [Latin *moles*, English translation *mass*] or density *ρ(r,t)* and intensity of *lumen* *ξ(r,t)* written as functions of the time *t* since the formation of the firmament and the radial distance *r* from the centre. (It is not evident from the text that Grosseteste envisaged that it took time for the *lumen* to propagate, but his description of the formation of the celestial spheres definitely describes a sequence of events, which we interpret in terms of time. We too assume that the lumen propagates instantly but that it takes time to generate the associated wave of matter.)

The equation describing the propagation of matter has three terms which describe firstly the geometric concentration of light in a spherical geometry, secondly the absorption of light as it passes through "imperfect" matter; and thirdly the generation of *lumen* by perfected matter. Grosseteste was clear that matter is moved by the force of light upon it, but by the same token, that this interaction reduces the intensity of *lumen* (and so its capacity to rarefy matter further). This effectively introduces the notion of opacity (although Grosseteste does not require a single Latin term for it). He does not possess a Newtonian conception of action and reaction, but it he is clear that the combined effect of the work done (assembly and rarefaction of matter) by *lumen* is, by the time it attains the sphere of the moon, such that "*this luminosity did not have sufficient power*". Indeed, the process of inwardly propagating concentration requires that the opacity of imperfect matter is significant for, as it turns out, if this were not the case, the geometric concentration of light generates a hole in the centre of the universe before the outer spheres have crystallised, or as Grosseteste would have put it, become perfected. A concept of incomplete transparency of the perfected spheres is hinted at in the text in relation to Grosseteste's discussion of diurnal motion. He argues that the lower spheres are lower in purity and therefore they receive diurnal motion in a weakened state 'since the lower a sphere the less pure and the more weak is the first corporeal light in it'. Following Grosseteste's identification of perfected matter as the source for lumen, we make the assumption that the intensity of the *lumen* is increased in proportion to the density of perfected matter. Thus the intensity of lumen pushing inwards increases as more and more of the outer parts of the universe reach the critical density.

We therefore write for the field of *lumen* the differential equation:

$$\frac{d\xi}{dr} = \kappa(\rho - \tau\rho_c)\xi - \gamma - \frac{\beta\xi}{r} \qquad [2]$$

The parameter $\kappa$ quantifies the notion of opacity we introduced above, and so gauges the degree of interaction of *lumen* with imperfect (or not-yet-perfected) matter, while $\gamma$ controls the degree of *lumen* generation in perfected matter (we allow the presence of this third term only in regions where matter is perfected). The term $\tau$ defines the transparency of 'perfected' matter, which we will see later is critical to production of a stable solution with a finite number of perfected shells. ($\tau=1$ corresponds to perfect transparency of perfected matter to *lumen*, lesser values permit even perfected matter to contribute to its absorption). Note that the first term, due to opacity, is positive, meaning that the lumen intensity $\xi$ decreases as the radius decreases, but the next two terms are negative, increasing the intensity of lumen as the radius decreases due to the geometry and additional emissivity respectively.

We have been careful to distinguish the density of perfected matter ($\rho_c$) from that of the total matter ($\rho$), the imperfect matter density being the difference of $\rho$ and $\rho_c$. Initially, we will assume that perfected matter is completely transparent (but will reconsider this assumption later). The last term describes the natural focussing of intensity in spherical geometry, where the dimensional parameter $\beta = 2$ in a three-dimensional cosmos. Allowing $\beta$ to change allows us to investigate other symmetries such as cylindrical ($\beta=1$) and planar ($\beta=0$).

Now we have an equation for the *lumen* intensity, we can consider its effect on the matter. The motion of matter is driven by the force of *lumen* but we will assume that Grosseteste envisaged that matter, in an Aristotelian framework, responds to this pressure instantaneously and does not feel any form of inertia. Therefore we assume that the velocity of the matter is simply proportional to the local intensity of *lumen*:

$$v(r,t) = \xi(r,t) \qquad [3]$$

The proportionality constant can be unity without loss of generality, since it only sets the timescale on which the system evolves. For inward motion $v$ is positive. The conservation of matter implicit in the text of *De luce* then implies that the velocity drives changes in the distribution of matter according to the continuity equation

$$\frac{\partial \rho}{\partial t} = \rho \frac{\partial v}{\partial r} + v \frac{\partial \rho}{\partial r} + \beta.\rho \frac{v}{r} \qquad [4]$$

Finally, we must set the criterion at which matter becomes perfected by the inward propagation of *lumen*. It is evident that simple use of the minimum density criterion $\rho_{c0}$ will result in a "snow-plough" effect with no discrete shells being formed. Closer reading of Grosseteste's text indicates that he recognised that for discrete shells to be perfected, the density of the inner shells must be greater than that of the outer ones. In order to explain diurnal motion, Grosseteste devised a model in which, '*the lower a sphere the less pure and more weak is the first corporeal light in it*' and '*…. that [light] of higher bodies is more spiritual and simple, while that of lower bodies is more corporeal and multiplied.*'. The physical and textual problems are largely removed by interpreting Grosseteste's model in terms of a quantised *critical ratio* of the densities of light (*lux* and *lumen*) to matter.

Of course it is entirely possible that the problems that we have identified with the fixed minimal density hypothesis are simply and strongly related, because Grosseteste understood that this would not naturally give rise to discontinuous spheres and so deliberately sought after a rule that would. He stated that a '*greater density arose*' in the inner parts of the second sphere, and furthermore light, '*which is simple in the first sphere is doubled in the second*'. The duplication [Latin: *duplicata*] and subsequent higher ratios of the nested spheres correspond to his articulation of a more subtle rule that would permit a 'quantization' of perfected matter into a finite number of spheres. We interpret Grosseteste's text describing the formation of the inner spheres as requiring that the density must exceed one of a series of quantised thresholds (i.e. a $\rho_c$ is in subsequent shells a factor 1, 2, 3, 4… greater than the lowest possible density $\rho_{c0}$) and that there must be sufficient combined *lux* and *lumen* to perfect the matter. Note that we could equivalently assume that the *lumen* must come in multiples of the intensity of *lux*, and that a critical ratio of matter to intensity was required. Our choice of discretisation criterion is also supported by a remarkable discussion in the early part of the text in which Grosseteste discusses how finite, numerical ratios may result from the quotients of infinite series (which he then links to infinite multiplication of light and matter).

**Numerical Solution**

In computing the solution of the above differential equations, we initially investigated the structure of the universes simulated when we varied the parameters $\alpha$, $\kappa$ and $\gamma$. As in modern cosmology, the initial state of the universe will profoundly influence its subsequent development and thus, the initial density distribution with radius following the universe's expansion is chosen as a key variable. Choice of the opacity parameter $\kappa$ reflects the critical importance of the strength of the interaction between *lumen* and mass, the important coupling constant in Grosseteste's cosmology. The parameter $\gamma$ too is related to the strength of the *lumen*-mass interaction and is also critical to the evolution of the universe from its initial condition.

When we solve numerically the mathematical formulation of the problem, we find that there is a complex interaction between the initial density profile, the intensity of the *lumen* (and the coupling of *lumen* to matter) and the opacity. It is very unlikely that this would have been apparent to Grosseteste. If the *lumen* intensity is high, or the initial density profile shallow, a large opacity is required to prevent sufficient *lumen* reaching the inner parts and triggering a disorderly "perfection" in which the spheres are discontinuous. For example, there are regions of the parameter space of high $\gamma$ and low $\alpha$ where our models generate fragments of the third sphere interspersed among the inner parts of the second. We could arbitrarily invent a new "rule" to prevent this, but that would not be supported by the text. We prefer simply to identify those regions of parameter space corresponding to cosmologies in which the opacity is sufficiently high that such a break-down is avoided.

The 3-D problem was solved numerically, the code being verified in one dimension, where an analytic solution is possible (see Appendix A). Because of the strong delta-function spike seen in the analytic solution, designing an appropriate numerical method was not straight-forward. Attempts to use an Eulerian mesh revealed that any simple scheme was too diffusive to accurately capture the narrow width of the delta-function spike. We found that a Lagrangian particle-based scheme was much more successful (see Appendix B), and follows the analytic derivation quite accurately.

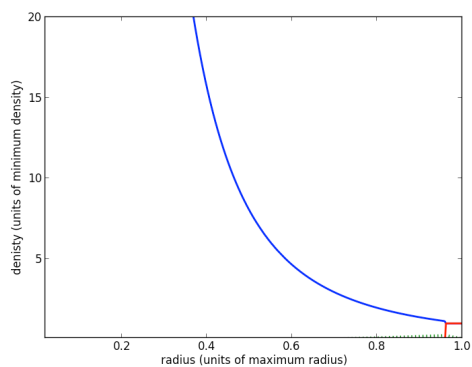

Fig 1(a)

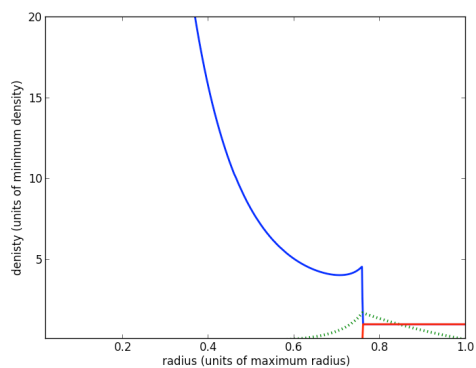

Fig 1(b)

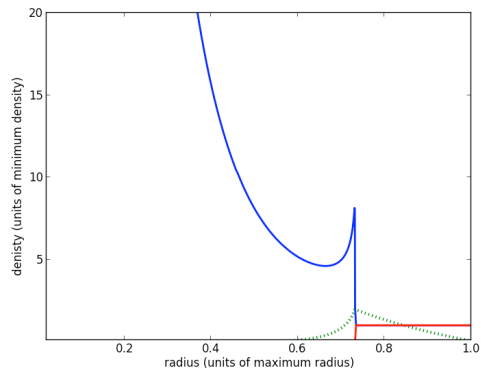

Fig 1(c)

Fig. 1 (a-c) Time sequence illustrating the formation of the outer sphere and the wave of matter pushed inward by the *lumen* that it generates. Imperfect matter is shown in blue, while the outer shell of perfected matter is in red. The dotted green line represents the *lumen* density. The model parameters are $\alpha = 3$, $\gamma = 5$, $\kappa = 5$ and $\tau = 1$. [An animated version of this figure is available at
http://www.durham.ac.uk/r.g.bower/
MedievalMultiverse]

The generic behaviour of the three dimensional model is illustrated in Fig. 1. The initial behaviour does not depend on the choice of parameters. After the actualisation of the outer shell, *lumen* propagates inwards sweeping up a wave of matter and leaving behind it a thick perfected sphere. Because of the opacity of matter, the *lumen* is more intense at the outer part of the wave. As the thickness of the outer sphere grows, the inwardly propagating *lumen* becomes more intense, speeding up the wave and making its crest sharper. The speed at which this sharpening occurs depends on the intensity of the *lumen* and the opacity of the matter. (Animated demonstrations are included in Electronic Supplementary Material.)

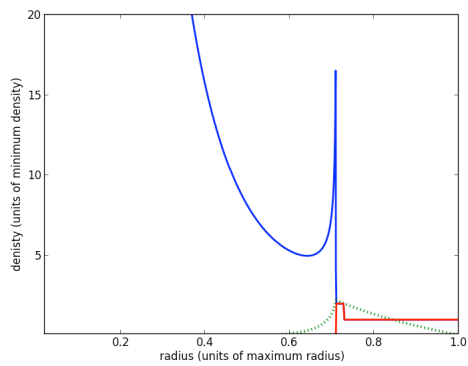

Fig 2(a)

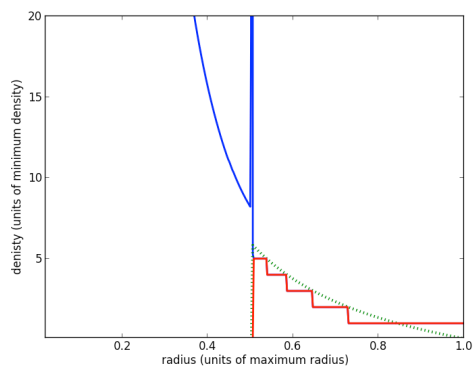

Fig 2(b)

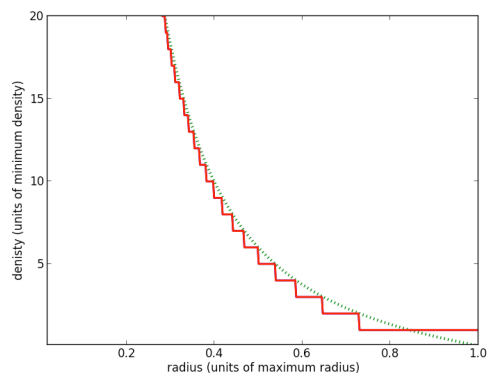

Fig 2(c)

Fig. 2 (a-c) Time sequence showing the subsequent formation of the outer sphere and the wave of matter pushed inward by the *lumen* that it generates, in a "well-behaved" region of parameter space. Imperfect matter is shown in blue; the outer shell of perfected matter is in red. Again, the green dotted line represents the *lumen* density. The model parameters are the same as Fig 1.

As the sweeping up of matter continues, the intensity of the *lumen* increases. Eventually the matter is dense enough and the *lumen* sufficiently intense that a second sphere, (i.e. in addition to the two-dimensional firmament,) can form. For suitable choice of parameters, the sweeping continues, eventually leading to the formation of the third sphere. As the process continues more and more spheres form; the initial steep profile of the matter density causing the shells to become ever more closely spaced (Fig. 2). The form of the universe predicted using this model is best visualised as a circularly symmetric two-dimensional pseudo-colour plot of the data (Fig. 3).

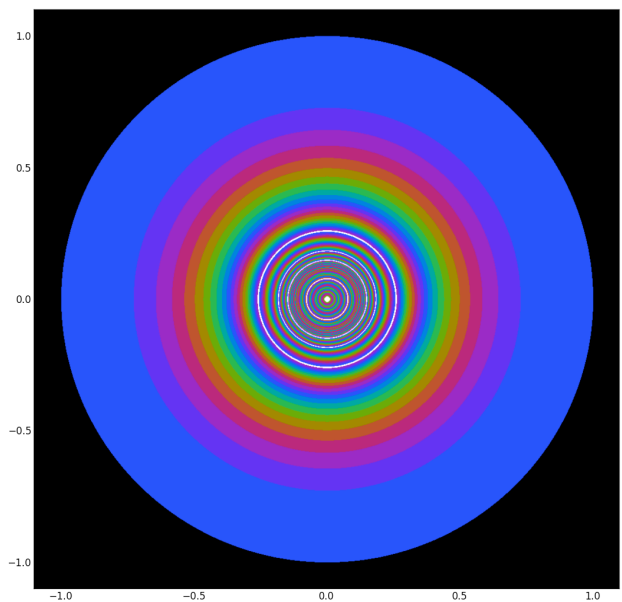

Fig. 3. A two-dimensional representation of the celestial spheres generated by Grosseteste's model (see Fig 2.) The model here assumes complete transparency of the perfected spheres. [An animated version of this figure is available at http://www.durham.ac.uk/r.g.bower/ MedievalMultiverse]

There is a major problem, however, as immediate examination of Fig. 3 shows that the model has generated too many spheres. In fact if we leave the computer model to run it generates an arbitrary number of spheres. This runs contrary to one of the fundamental ideas of Grosseteste's universe: that the *lumen* intensity will eventually be insufficient to crystallise the inner-most shells, leaving them imperfect and only partially separated. To be faithful to the significance invested by Grosseteste, as all medieval authors, in the perfected substance of the spheres, we should differentiate between the properties of *lumen*-mass interaction in the case of perfected and unperfected matter. We have termed the absorption of *lumen* in its passing through, and work on, unperfected mass as *opacity* – defined by the parameter $\kappa$. Additionally, there is the possibility of absorption of *lumen* by already-perfected matter. This property we might term *transparency*, and use the symbol $\tau$ in order to be clear that it is a different physical issue. *De luce* permits interpretations of both complete transparency *($\tau=1$)* and incomplete transparency *($\tau<1$)*. (In all cases we assume that the first sphere is completely transparent). In our numerical exploration, we have found that it is impossible to build a cosmos with a finite number of spheres if the perfected spheres are completely transparent. As the outer spheres are perfected, the intensity of *lumen* in the inner parts increases to arbitrarily high values, allowing the perfection of matter at ever higher densities. This is not how Grosseteste imagined the earth and the other sublunar spheres were formed. Eventually, he argued, the *lumen* is too weak compared to the matter density that perfection is not possible and only a partial separation of earth, water, fire and air occurs. This is how he explains the phenomenon of weight of water and earth, and the existence of radial motion.

In order to reproduce this effect in our model, we must assume that the "actualised" outer spheres are not completely transparent. This is done by setting a value for the parameter $\tau$ of less than unity and it does enable us to find situations under which the *lumen* intensity shows an overall decrease with decreasing radius. Such conditions can lead to the intensity dropping sufficiently so that the perfection process stalls. This is the type of universe in which Grosseteste imagines we live. Although a non-unity value of $\tau$ is required for simulation of a small number of perfected spheres, we have limited the reduction in transparency so as not to conflict with 13[th] century ideas and remain faithful to Grosseteste's text.

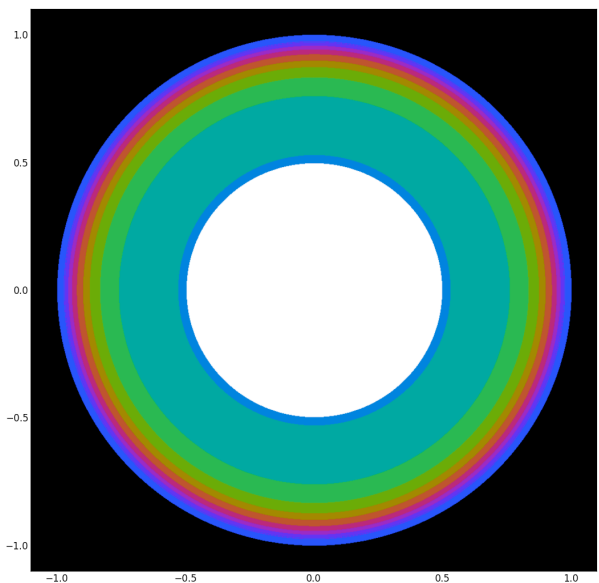

Fig. 4. Two-dimensional representation of a convergent simulation of a universe with nine perfected spheres. The central sphere contains imperfect matter as the perfection process stalls as the model starts to generate a tenth sphere. Model parameters are $\alpha = 30$, $\gamma = 55$, $\kappa = 2.5$ and $\tau = 0.6$. By adjusting the other parameters, similar results can be obtained with $\tau$ as large as 0.9.
[An animated version of this figure is available at http://www.durham.ac.uk/r.g.bower/MedievalMultiverse]

It is necessary that the *lumen* from the outer parts is absorbed faster than the intensity increases but even so, a careful balance of the model parameters is required to ensure that the process creates more than one or two spheres before the *lumen* intensity becomes insufficient. An example of a cosmos with nine actualised spheres is shown below in Fig. 4. At a radius of 0.6, the absorption of *lumen* from the other spheres is such that the *lumen* intensity actually declines as the innermost shell is perfected and the process of actualisation stalls. However long we run the calculation, the inner core of matter will never be perfected. We note that the radius of the four sub-lunar spheres, (which Grosseteste considers as a single unperfected sphere,) is much too large in comparison with the other celestial spheres. The universe depicted does not correspond to the geometry envisaged by medieval philosophers.

The required condition for the *lumen* intensity to show an overall decrease with decreasing radius is

$$\kappa((1-\tau)\rho_c) > \frac{2}{r} + \frac{\gamma}{\xi} \qquad [5]$$

Note that the left hand side is a function of the perfected density, not the initial density, and that it is larger as more shells are formed. On the other hand, it becomes increasingly difficult to satisfy this condition as r decreases, so that the process can only stall if the left-hand term is sufficiently large immediately a new sphere starts to form.

Since the formation of a new sphere must be limited by the *lumen* intensity (otherwise the spheres become disordered), we have $\xi=\rho_c$. For sufficiently, large $\xi$ and small *r*, the first term dominates the right hand side and we can see that there is a minimum radius at which the stall can occur

$$r_{stall} > \frac{2}{\kappa(1-\tau)\rho_c} \qquad [6]$$

In order to have the process stall after generating a large number of spheres, this condition needs to be satisfied only when $\rho_c$ is large. As the opacity varies, the balance of these terms shifts, generating bands of stalled models with different numbers of perfected spheres.

**A medieval multiverse**

Stable universes with a finite number of spheres are very much the exception. The *lumen* intensity and the opacity both need to be very high. If the *lumen* intensity is low, the spheres form at small radius where the geometric concentration is dominant. If the opacity of the matter is too small, the formation of the second sphere begins before sufficient mass has been swept up and then the inward propagation of *lumen* results in non-monotonic density with radius. The result is a strange cosmos in which the spheres are disordered and mixed up with each other.

A familiar concept in modern cosmology is to consider the universe we live in to be just one of many possible universes, each individual universe differing in the value of its fundamental parameters. We can then attempt to determine if the universe we live in is in some way special, or the result of taking a random selection from many possibilities. We can apply the same logic to Grosseteste's universe, and quickly discover that the organised Aristotelian universe, with nine perfected spheres and a tenth sphere of only partially separated elements requires a very special combination of the fundamental parameters.

Within the ($\alpha, \kappa, \gamma, \tau$) parameter space, it is possible to generate cosmologies (Appendix C) with any number of spheres, encompassing both zero and an infinite number. Only in a very limited part of parameter space are universes with over six perfected shells predicted. Examination of the solutions in ($\alpha, \kappa, \gamma$) space at specific values of $\tau$ reveals that this occurs when both the initial density gradient and *lumen* generation parameters, $\alpha$ and $\gamma$ are very high.

The nature of universes predicted by Grosseteste's model as a function of the ($\kappa, \gamma, \tau$) parameter space, for $\alpha = 35$, is displayed in Fig. 5. (An animated version is included in the Electronic Supplementary Material.) Here we see that the model does generate a ten-sphere universe but only in a very limited region of parameter space. As indicated in the discussion above, the *lumen* generation intensity parameter $\gamma$ has to be very high. Parameters relating to opacity/transparency ($\kappa$ and $\tau$) and interaction of mass and light ($\gamma$) have to be finely tuned in order to yield the type of universe that Grosseteste envisaged ($\alpha$ must also be sufficiently large). Although ten-sphere solutions can be found for a wide range of $\kappa$ and $\tau$ values, the manifold is extremely thin.

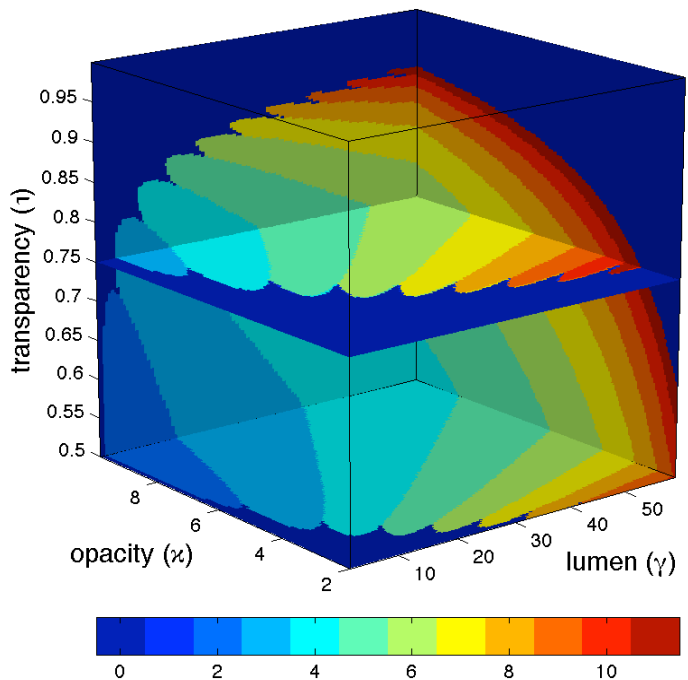

Fig. 5. The "medieval multiverse" showing the regions of stability in opacity $\kappa$, *lumen* intensity $\gamma$ and transparency $\tau$ space within which plausible universes exist. The colours correspond to the different numbers of spheres created by the model. [An animated version of this figure is available at http://www.durham.ac.uk/r.g.bower/MedievalMultiverse]

Following the same logic as modern cosmologists, we are forced to conclude that some additional physical law is at work that singles out points in the parameter space corresponding to the universe we inhabit. A contemporary example would be the "anthropic principle". Grosseteste resorted to a Pythagorean numerology in the closing section of the *De luce*, arguing that the number 10 has special qualities of perfection and thus there were nine spheres, including an unperfected sphere of the elements, below the firmament; ten in total. Although Grosseteste specified his cosmology remarkably closely – enough to enable us to map his physical axioms onto computable mathematical statements, this analysis shows that even this degree of precision leaves open (at least) a three-parameter family of universes. This motivates a further question of his thinking – to what extent did he contemplate the possibility of different universes? If he did, did he imagine their existence as equivalent to the one he inhabited, following Anselm [11], who *Cur Deus homo I.21,* debates the theological implications of multiple worlds similar to our own in the context of sin? Or did he regard alternatives as simply potential? Debate on the subject existed throughout the 13[th] century over whether there might be multiple actual universes. Aristotle's *De caelo* [12], which postulates an eternal and singular universe, had been available to western authors since the 12[th] century. Although the answer in Grosseteste's day was negative, the notion of multiple universes became a subject closely bound up with the question of divine omnipotence. Article 34 of Bishop Stephen Tempier's famous Condemnations of 1277, explicitly condemned the notion that God could not produce more than one world [13]. We cannot know Grosseteste's view, but the computer simulations have revealed a fascinating depth to his model of which he was certainly unaware. In particular, the sensitivity to initial conditions resonates with contemporary cosmological discussion and reveals a subtlety of the medieval model which historians of science could never have deduced from the text alone. The results provide a striking example of the benefits of our methodology of collaborative reading of texts by a team which contains medievalists, linguists and scientists. The approach is uniquely different from traditional models of cooperation between humanities scholars and scientists, where each group operates strictly within its own discipline to address different questions of common interest. In our project, the groups work together on a single question, approaching it from different conceptual frameworks.

**Appendix A**

**Comparison of analytical and numerical solutions in one dimension**

Let us consider solution in one dimension and assume that the initial density is constant ($=\rho_0$). We ignore the crystallisation of matter at non-zero density, and we assume that lumen is generated only at the outer boundary. These simplifications allow us to assert the lumen acting on an element of the matter depends only on its initial position, $r_0$.

$$\frac{d\xi}{dr_0} = \kappa \rho_0 \xi \qquad [A1]$$

For convenience, we here adopt a coordinate system in which $r > 0$ and the initial lumen intensity, $\xi_0$, is incident on the inner edge of the calculation, at $r = 0$. (This is different to the coordinate system used in solution of the spherical problem, but the two coordinate systems can be simply related by the mapping $r \rightarrow 1-r$.)

We take a Lagrangian approach, tagging each matter element by its initial position. As the matter elements move in response to the *lumen*, they retain their original ordering, and thus the *lumen* acting on them remains constant. The velocity of an element, *v*, therefore remains constant

$$v = \xi_0 e^{-\kappa \rho_0 r_0} \qquad [A2]$$

If the particles do not overtake one another, their position at time *t* is simply

$$r(t) = r_0 + vt \qquad [A3]$$

but the situation is complicated because the speed of the particles with smaller $r_0$ will be largest. Eventually, the particles at small r will sweep up the particles in front of them forming a delta function spike in the density distribution. The delta function does not disperse, nor the initially faster particles overtake, because all particles in the delta function distribution experience the same

*lumen* field, so travel with identical velocity. The position of the density spike is determined by the distance travelled by the particle that starts at $r_0 = 0$. To determine the mass of the spike, we need to identify the *initial* radius of the last particles to be swept up, $r_2$. If $r_s$ is the position of the spike at time $t$,

$$r_s(t) = \xi_0 t = r_2 + \xi_0 t e^{-\kappa p_0 r_2} \qquad [A4]$$

This allows us to solve for $r_2$ by tabulating it as a function of time:

$$\xi_0 t = \frac{r_2}{1 - e^{-\kappa p_0 r_2}} \qquad [A5]$$

The mass of the swept up spike is $\rho_0 r_2$.

Particles initially more distant from the origin than $r_2$ will be compressed increasing the density at position $r$. This creates a tail to the delta function.

$$\rho_0(r) = \rho_0 \frac{\partial r_0}{\partial r} = \frac{\rho_0}{1 - \kappa p_0 \xi_0 t e^{-\kappa p_0 r_0}} = \frac{\rho_0}{1 - \kappa p_0 (r - r_0)} \qquad [A6]$$

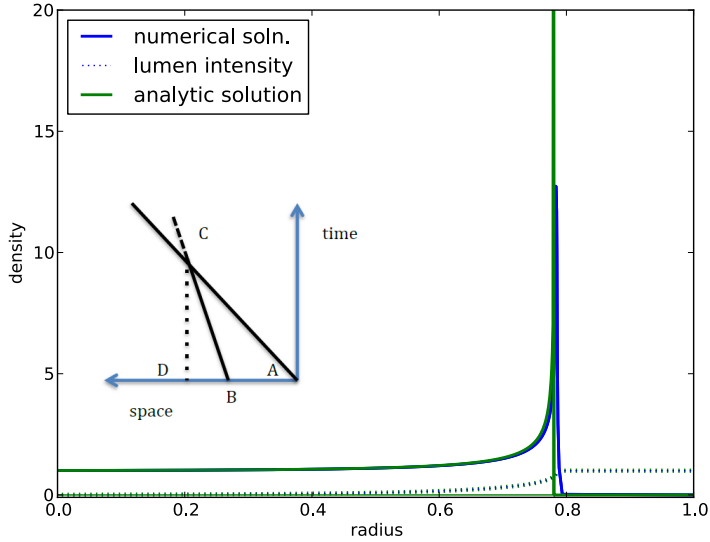

Fig A1. The solution to the "Grosseteste equations" in the simplified case described above. The solid green line shows the analytic solution, while the solid blue line shows the numerical solution. The lumen intensity is shown as a dotted line. While the delta function spike is infinitely thin in the analytic model, it has finite width in the numerical solution but the numerical solution accurately tracks it and the tail that forms ahead of it. The model shown has $\kappa = 3$ and a uniform initial density, $\rho_0 = 1.1$. The lumen intensity at the left boundary is $\xi_0 = 1$. Inset: A geometric representation of the formation of the mass delta function. In this space-time representation, the slope of the lines represents particle velocities. The particle with initial position B is overtaken by A at point C. All the mass between A and B is swept up into a delta function spike.

An example solution is shown in Fig A1. (In order to simplify comparison with the other figures in the paper, we plot *1-r* on the x axis so that the front of swept up matter moves from right to left.) The inset shows geometrically how the delta function arises. We assume that lumen is incident from the left, moving "particles" to the right. The particle velocity is represented by the slope of the line in the diagram. Because of the opacity, particles closer to the source of lumen move most quickly, and particle A will overtake particle B at point C. At this time the location of the spike will be point D (ADC is a right angle) and the mass of the swept-up shell is given by the length AB. In order to complete the problem we need only to specify how the angle CBD (i.e. the velocity of B) depends on the location of B. The reason that the faster particles do not actually overtake the slower, but instead cause the delta-function pile-up, is that when particles are swept up by the delta function they feel the full force of non-absorbed lumen for the first time so instantaneously accelerate to the velocity of the delta function.

## Appendix B

**Numerical Solution**

Because of the strong delta-function spike that is seen in the analytic solution, designing an appropriate numerical method is not straight-forward. We initially attempted to solve the problem using an Eulerian mesh. However, we found that any simple scheme was too diffusive to accurately capture the narrow width of the delta-function spike. Appropriate description of the perfection conditions then becomes problematic.

We found that a Lagrangian particle based scheme was much more successful, and follows the analytic derivation quite accurately. At the start of each time step, we compute the local density from the particle separations and solve for the lumen intensity acting on each particle. We use this to derive particle velocities, and to determine the density at which a particle will become perfected. This is either the minimum matter density ($\rho = 1$), or the integer density just lower than the smaller of the lumen intensity or the matter density. In practice, the perfection must be limited by the lumen intensity if the formation of spheres is to proceed in an ordered manner. Particles that have already been perfected are given zero velocity. A time-step is then chosen so that the most fast moving particles only move a fraction of the spatial resolution. Particles are then moved in order, starting from the outermost particle (which has not been perfected) and working inwards until the velocity is negligible. For each particle, a trial step is taken based on the velocity and global time-step. If this results in the new density of the particle falling below its perfection level, we sub-cycle in order to set the particle position from the perfection level. In this way we avoid over estimating particle movements and missing the correct level for the phase change. Once all the particles have been moved, we begin a new step by re-computing densities and phase transition levels. The initial spacing and densities of the particles are set to sample the initial density profile. We have experimented with using both fixed spacing but strongly varying initial densities, and by generating the initial profile using an adaptation of the initial spacing. Given the very strong density contrasts in the problem we find that the best technique is to combine both approaches, although they do give equivalent answers. We quote the resolution in terms of a uniform spacing.

We have verified the code by comparing with the analytic solution discussed in Appendix A (for the simplified problem), and by comparing the solution calculated with different spatial resolution (for the full problem). We adopt a Courant factor of 0.1 in order to set the global time-step. The code converges extremely well, even in complex cases. Fig. B1 illustrates an example where the perfection process eventually stalls because of the limited transparency of the perfected spheres. The three lines illustrate different initial particle spacing, ranging from 1% to 0.1% of the outer radius. The solutions are essentially identical. Typically we will run the code with an initial spacing of 0.2% in order to capture the fine details of the perfection process in the inner parts.

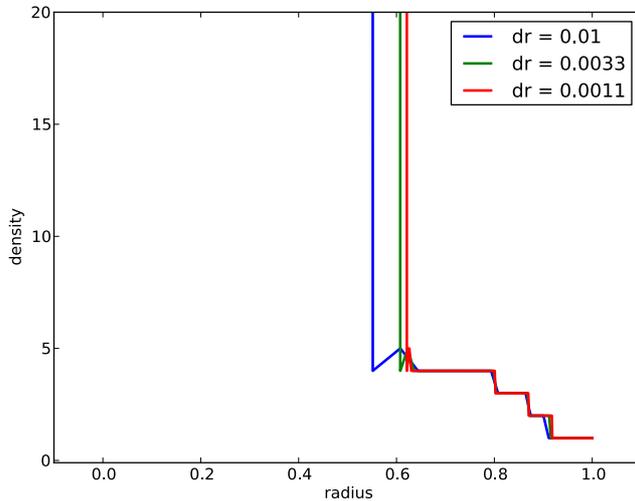

Fig B1. Illustration of the effect of different initial particle spacing on the numerical solution. The different (coloured) curves indicate the initial spacing of particles. The parameters used are $\kappa=5$, $\gamma=20$, $\alpha=10$.

# Appendix C

**A technique for rapid solution**

While the numerical code we have built is fast enough for solving the equations for a few sets of parameters, it is not fast enough to explore the parameter space of the multiverse. Fortunately, if the initial density profile is sufficiently steep, the behaviour of the solution does not depend on the detail of the motion of the matter since the level at which perfection takes place is always set by the lumen intensity. Under this assumption the final perfection of the matter can be determined from the lumen equation alone. We show below that this approach is extremely accurate.

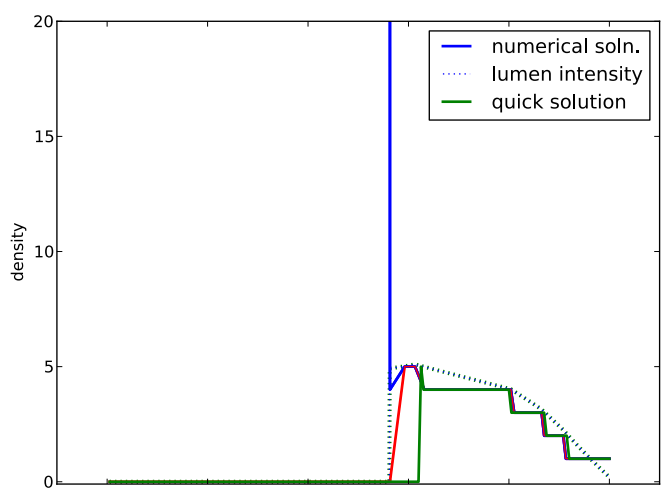

Fig C1. A comparison of the full numerical solution (blue and red) and the quick solution method (green) that solves directly for the lumen intensity. The parameters used are $\kappa = 5, \gamma = 20$ (as in Fig B1). The initial density slope parameter $\alpha$ is 10, but this is not relevant to the quick solution.

In order to solve for the perfected density, we first compute the lumen intensity on a uniformly spaced radial grid. We then determine the perfected density of the last unperfected radius by taking the integer part of the lumen; we are thus assuming that lumen not the matter density sets the perfection level. Perfecting this shell requires us to re-compute the lumen distribution, but we can then determine the level at which the next shell is perfected, assuming that sufficient matter has been swept up to permit completion at this level. In this way we can quickly compute the final configuration, and identify whether the perfection process will stall, without following the matter flow. C1 shows an example of the rapid solution, comparing it to the solution obtained with the full numerical method. We used an initial spacing of 5% for the numerical solution. Experiments with finer spacing show that the quick approach is the more accurate.

**Acknowledgements**

Financial support from the Arts and Humanities Research Council and the Welcome Trust is gratefully acknowledged, as is the support of the Science and Technology Facilities Council [grant number ST/F001166/1] . The contributions of Greti Dinkova-Bruun, Faith Wallis, Philip Anderson, Michael Huxtable, Pietro Rossi and Till Sawala during collaborative-reading workshops have been of enormous value to the outcomes of this interdisciplinary project. The data used in this project can be obtained by e-mailing r.g.bower@dur.ac.uk.